# Sedimentation of Reversibly Interacting Macromolecules with Changes in Fluorescence Quantum Yield

S.K. Chaturvedi, H. Zhao, P. Schuck*

*schuckp@mail.nih.gov

## Abstract

Sedimentation velocity analytical ultracentrifugation with fluorescence detection has emerged as a powerful method for the study of interacting systems of macromolecules. It combines picomolar sensitivity with high hydrodynamic resolution, and can be carried out with photoswitchable fluorophores for multi-component discrimination, to determine the stoichiometry, affinity, and shape of macromolecular complexes with dissociation equilibrium constants from picomolar to micromolar. A popular approach for data interpretation is the determination of the binding affinity by isotherms of weight-average sedimentation coefficients $s_w$. A prevailing dogma in sedimentation analysis is that the weight-average sedimentation coefficient from the transport method corresponds to the signal- and population-weighted average of all species. We show that this does not always hold true for systems that exhibit significant signal changes with complex formation – properties that may be readily encountered in practice, e.g., from a change in fluorescence quantum yield. Coupled transport in the reaction boundary of rapidly reversible systems can make significant contributions to the observed migration in a way that cannot be accounted for in the standard population-based average. Effective particle theory provides a simple physical picture for the reaction-coupled migration process. On this basis we develop a more general binding model that converges to the well-known form of $s_w$ with constant signals, but can account simultaneously for hydrodynamic co-transport in the presence of changes in fluorescence quantum yield. We believe this will be useful when studying interacting systems exhibiting fluorescence quenching, enhancement or Förster resonance energy transfer with transport methods.

**Length Estimate:** 33,783 characters (with spaces) and 7 spaces for figures = 7.4 pages

**Introduction**

Sedimentation velocity analytical ultracentrifugation (SV) is a powerful technique for the characterization of protein interactions (1–5). Unique among classical transport methods is the mass-based driving force, which causes strongly size-dependent hydrodynamic migration of molecules free in solution, such that sedimentation data from various real-time optical detection systems can be interpreted on the basis of first principles of chemical reactions and hydrodynamics. During the last decade, SV has rapidly evolved with new and increasingly sophisticated computational analyses approaches (6–12), new theoretical relationships and hydrodynamic methods (13–18), and new experimental platforms, calibration and detection systems (19–25). They are supporting an increasing variety of applications, extending SV from classical applications of protein size-and-shape measurements and studies of protein interactions to areas such as carbohydrate interactions (26, 27), protein self-assembly and aggregation (28–31), membrane proteins and nanodiscs (32–34), nanoparticles (35–40), and biotechnology (41–43).

In particular, a significant extension of the range of application of SV was afforded by the recent introduction of a commercial fluorescence detection system (FDS-SV) (19, 44, 45). After accounting for a characteristic signal structure imposed by the optical configuration (46), it offers highly quantitative representation of the sedimentation process (46, 47), and allows the detection of low picomolar concentrations of common fluorophores (32, 48). This has permitted studies of labeled proteins in complex fluids such as serum (49, 50) and cell lysates (29, 51), and diverse studies of high-affinity interacting systems (48, 52–55) with $K_d$ as low as 20 pM (48).

An important consideration when working with fluorescence detection in SV is the potential for photo-physical effects that may impact the detection efficiency. We have shown previously that common fluorophores such as fluorescein and enhanced green fluorescent protein (EGFP) do not exhibit measurable irreversible photobleaching in the relatively weak and transient sample illumination in the spinning rotor during SV (56). By contrast, reversibly photoswitchable fluorescent proteins have a much higher quantum yield of photo-conversion, and exhibit temporally modulated sedimentation signals (56). These can be easily quantitatively accounted for and folded into the sedimentation analysis, and may even be exploited as specific tags to enable monochromatic multi-component analysis for FDS-SV (57). Another category of photophysical effects involves energy transfer to other molecules, which may lead to specific quenching of signals of protein complexes, thereby potentially offering additional proximity-based information in interacting systems. However, so far this has not been experimentally explored, and the impact of static or dynamic quenching --- or more generally changes in fluorescence quantum yield on complex formation --- on the sedimentation analysis has not been examined.

In the present work, we consider theoretically the sedimentation analysis of a system where fluorescent free species co-exist with faster-sedimenting complexes with altered fluorescence quantum yield, under conditions of rapid interconversion relative to the time-scale of sedimentation. Given the prevalence of static or dynamic quenching upon complex formation (58–61), e.g., measured in benchtop fluorescence binding experiments, or opposite, the possible enhancement of fluorescence upon binding (62, 63), as exploited, for example, in fluorescence complementation sensors (64, 65), and the possibility of protein complexes --- even with high-affinity interactions (66) --- to exhibit off-rate constants in excess of 10$^{-2}$/sec, this is a highly relevant scenario. We show that in this case the signal-weighted average sedimentation coefficients $s_w$ from the transport method --- the most basic analysis of interacting systems --- cannot be simply understood as a population averaged transport, as widely assumed. Trivial corrections merely accounting for different species' signal contributions are possible for the case of slow chemical exchange, but fail when sedimentation of different species is coupled. In order to understand the mechanism of sedimentation of rapidly reversible interacting systems, we have previously developed the effective particle model as a physically intuitive picture of the coupled sedimentation process (14). It quantitatively describes the experimentally observed boundary patterns as a function of loading concentrations, which is comprised of an undisturbed boundary sedimenting at the rate of one of the free species, and the reaction boundary with concentration-dependent velocity (14, 67). We demonstrate here that the effective particle model also provides a simple framework for the interpretation of $s_w$ isotherms in the presence of binding-induced changes in fluorescence quantum yield.

## Theory

### Sedimentation of Interacting Systems

As a simple model system we consider a smaller component **A** binding a larger component **B** to form a bimolecular complex **AB**, obeying the mass action law K=[AB]/[A][B]. In the centrifugal field, the evolution of the local concentration of each macromolecular species in solution, $\chi_i(r,t)$, is given by the Lamm equation

$$\frac{\partial \chi_i(r,t)}{\partial t} = -\frac{1}{r}\frac{\partial}{\partial r}\left(\chi_i s_i \omega^2 r^2 - D_i \frac{\partial \chi_i}{\partial r} r\right) + q_i \qquad \text{(Eq. 1)}$$

with the time denoted as $t$, distance from the center of rotation $r$, sedimentation coefficient $s$ and diffusion coefficients $D$, respectively, and rotor speed $\omega$ (68). The terms in parenthesis are sedimentation and diffusion fluxes, respectively, followed by chemical reaction fluxes $q_i$ that couple sedimentation of all species (68). The evolution of the total signal measured in the SV experiment is the sum of signals from all species

$$a(r,t) = \sum \varepsilon_i \chi_i(r,t)$$
$$= \varepsilon_A \chi_A(r,t) + \varepsilon_B \chi_B(r,t) + (\varepsilon_A + \varepsilon_B + \Delta\varepsilon)\chi_{AB}(r,t)$$

(Eq. 2)

Here we have denoted the component signal coefficients as $\varepsilon_A$ and $\varepsilon_B$ (absorbing optical pathlengths), respectively, and allowed for a deviation of the complex signal from the sum of the components' signals, expressed as $\Delta\varepsilon$. $\Delta\varepsilon$ may be the result of static or dynamic quenching, fluorescence enhancement, or generally any process changing the fluorescence quantum yield when the fluorophores are the complex. The coupled Lamm equation was solved using the software SEDPHAT as described in (69) (http://sedfitsedphat.nibib.nih.gov/software).

### *The Signal-Weighted Sedimentation Coefficient*

Based on the transport method (5, 70, 71), the signal-weighted average sedimentation coefficient is defined by the change in the mass balance of macromolecules that remain in the solution column between meniscus $m$ and a plateau radius $r_p$ (2, 71). With the measured radius- and time-dependent signal denoted as $a(r,t)$, $s_w$ can be defined as (5)

$$s_w =: \lim_{t \to 0}\left[ \frac{1}{a(r_p,t)\omega^2 r_p^2}\left( -\frac{d}{dt}\int_m^{r_p} a(r,t)r\,dr \right) \right]$$

(Eq. 3)

. (In slight deviation from previous definitions, the limit to zero time is essential for removing the time-dependence arising from differential radial dilution in heterogeneous systems (5), and to achieve consistency with the results from weighted integration of sedimentation coefficient distributions; see below).

Previously, it has traditionally been considered self-evident that $s_w$ is identical to a signal- and population-weighted average of all species in solution (2, 67, 72–74) (if considering signal coefficients at all). This dogmatic approach leads to the form

$$s_w^0 = \frac{\sum \varepsilon_i c_i s_i}{\sum \varepsilon_i c_i} = \frac{\varepsilon_A c_A s_A + \varepsilon_B c_B s_B + (\varepsilon_A + \varepsilon_B + \Delta\varepsilon)c_{AB}s_{AB}}{\varepsilon_A c_{A,tot} + \varepsilon_B c_{B,tot} + \Delta\varepsilon c_{AB}}$$

(Eq. 4)

with $c_i$ denoting species loading concentrations, and $c_{tot}$ denoting total component loading concentrations.

However, it is possible to arrive at the $s_w$ isotherm more rigorously by derivation from the definition Eq. 3. To this end, we insert the signal-coefficient weighted Lamm equation solutions (Eq. 1) into Eq. 3

$$s_w = \lim_{t \to 0} \left[ -\frac{1}{a(r_p,t)\omega^2 r_p^2} \left( \int_m^{r_p} \sum_i \varepsilon_i \frac{d\chi_i(r,t)}{dt} r dr \right) \right] \qquad \text{(Eq. 5)}$$

and expand the chemical reaction and migration terms

$$s_w = \lim_{t \to 0} \left[ -\frac{1}{a(r_p,t)\omega^2 r_p^2} \left( \int_m^{r_p} \sum_i \varepsilon_i q_i r dr \right) \right]$$
$$+ \lim_{t \to 0} \left[ \frac{1}{a(r_p,t)\omega^2 r_p^2} \left( \int_m^{r_p} \sum_i \varepsilon_i \left[ \frac{\partial}{\partial r} \left( \chi_i s_i \omega^2 r^2 - D_i \frac{\partial \chi_i}{\partial r} r \right) \right] dr \right) \right] \qquad \text{(Eq. 6)}$$

For the second term, as shown in (5, 75), we can execute the now trivial integral, and exploit the fact that the total flux vanishes at the meniscus, and that the gradients vanish in the plateaus. In the limit of zero time, the plateau concentration approaches the loading concentration, and therefore the second term equals $s_w^0$ of Eq. 4:

$$s_w = s_w^0 + \lim_{t \to 0} \left[ -\frac{1}{a(r_p,t)\omega^2 r_p^2} \left( \int_m^{r_p} \sum_i \varepsilon_i q_i r dr \right) \right] \qquad \text{(Eq. 7)}$$

Thus, we find that the signal-weighted sedimentation coefficient is identical to the conventionally population-averaged $s_w^0$ *only* if $\sum \varepsilon_i q_i$ vanishes, i.e., signals are conserved in the interaction. For the bimolecular model system, due to mass conservation, the reaction fluxes are of equal magnitude ($q_A = q_B = -q_{AB} = q$) and therefore $s_w$ takes the usual form in the absence of signal changes in the complex ($s_w = s_w^0$ if $\Delta \varepsilon = 0$). Alternatively, the naïve $s_w^0$ holds true if $q = 0$, i.e., when there is no reaction on time-scale of sedimentation. We note if there are signal changes for slow reactions ($\Delta \varepsilon \neq 0$ and $q = 0$), then it solely affects the magnitude of complex signal, as if there was a different binding constant

$$K_{app} = \frac{\varepsilon_A + \varepsilon_B + \Delta \varepsilon}{\varepsilon_A + \varepsilon_B} K \qquad \text{(Eq. 8)}$$

measured under neglect of signal changes. However, a completely different situation occurs when signal changes are associated with coupled transport. Then the second term of Eq. 7 makes significant and non-trivial contributions that require knowledge of the Lamm equation solutions. To this end, we turn to approximate solutions of Eq. 1 in the absence of diffusion.

## *Boundary Patterns and Effective Particle Theory (EPT)*

Even though the Lamm PDE Eq. 1 allows the precise prediction of macromolecular migration given particular set of transport and reaction parameters and loading concentrations, as a statement of local fluxes and mass balance it does not directly provide physical insight in the mechanisms of coupled transport or principles governing the resulting boundary patterns. However, much can be deduced about the nature of the coupled sedimentation process from approximations in linear geometry with constant force and negligible diffusion, an approximation introduced first by Gilbert and Jenkins in their seminal theory (GJT) (76), and later also applied in effective particle theory (EPT). While GJT was concerned with polydispersity of velocities in the 'asymptotic boundaries' at infinite time, the focus of EPT is on quantities that can be measured easily in experiments: the average amplitude and sedimentation coefficients of the ensuing sedimentation boundaries of the reacting system as a function of loading concentrations (14). EPT exploits Heaviside step-functions to model the diffusion-free boundaries, which collapses the Lamm PDE to a set of simple algebraic relationships highlighting the underlying physics; for details see (14). EPT naturally explains many otherwise seemingly counter-intuitive features that have been experimentally observed (67, 68).

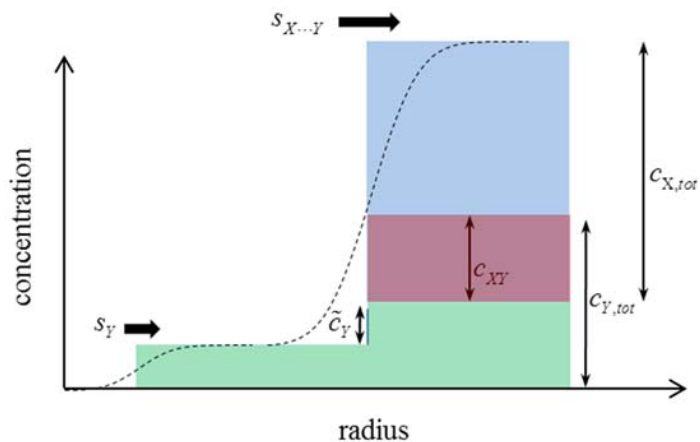

Figure 1: Schematic structure of the sedimentation pattern of a rapidly reversible two-component interacting system. The solution column extends from left to right, and the height of the different colored areas reflects the local concentration of different species, including free **Y** (green), complex **XY** (red), and free **X** (blue). In the configuration depicted, the undisturbed boundary only contains a fraction of the total of **Y** at a concentration $c_{Y,tot} - c_{XY} - \tilde{c}_Y$ and migrates with the velocity $s_Y$, which is smaller than the velocity of the reaction boundary $s_{X\cdots Y}$. In the reaction boundary, all species of component **X** are present, including the complex and the co-sedimenting fraction of **Y** at $\tilde{c}_Y$. The magnitude of $\tilde{c}_Y$ is such that the time-average velocity of all co-sedimenting **Y** (comprising $\tilde{c}_Y$ and $c_{XY}$) equals the time-average velocity of all **X** (comprising $c_X$ and $c_{XY}$). The concentration of complex $c_{XY}$ is governed by mass action law for the given total concentrations $c_{X,tot}$ and $c_{Y,tot}$. In effective particle theory, the diffusion-broadened boundaries (dotted line) are approximated as step-functions that migrate with the velocities and amplitudes such that they satisfy the coupled system of Lamm equations (14).

Briefly, rapidly interacting two-component systems sediment according to the following principles (Fig. 1) (14): To satisfy mass action law at all positions within the solution column at all times, there can only be one 'reaction boundary' containing all components. It completely engulfs one of the components, for which we use the nomenclature $\mathbf{X}$, comprising both its free form (at the concentration of free $c_X$) as well as all of the complex (at concentration $c_{XY}$). Additionally, it also co-transports as much of the other component $\mathbf{Y}$ (at free concentration $\tilde{c}_Y$) such that the time-average velocity of both boundary constituents will match ($s_{X\cdots Y}$). We reemphasize the important fact that the reaction boundary will contain free species of all components, though in different proportions to compensate for their differing $s$-value when free. The excess of $\mathbf{Y}$ over what can be transported along with the reaction boundary, if any, will sediment as free species in the 'undisturbed boundary'. The coupled transport process can be visualized as animated cartoons generated by the Effective Particle Explorer tool in SEDPHAT for given loading concentrations and sedimentation velocities.

The selection which component is $\mathbf{X}$ and $\mathbf{Y}$ is determined by a 'phase transition' line $c_{B,tot}^*(c_{A,tot})$ separating the two cases. At the transition line, the undisturbed boundary disappears and the system sediments in a single boundary. Quantitatively, using a terminology of $\mathbf{A}$ and $\mathbf{B}$ such that $s_A \leq s_B$, EPT predicts the phase transition for a simple bimolecular reaction to occur at

$$c_{B,tot}^*(c_{A,tot}) = c_{A,tot} + \frac{(s_B - s_A)}{2K(s_{AB} - s_B)}\left(1 + \sqrt{1 + \frac{4c_{A,tot}K(s_{AB} - s_B)}{(s_{AB} - s_A)}}\right) \qquad \text{(Eq. 9)}$$

, such that $\mathbf{X}$ becomes $\mathbf{A}$ for $c_{B,tot} > c_{B,tot}^*$, and $\mathbf{X}$ becomes $\mathbf{B}$ otherwise.

EPT leads to the velocity of the reaction boundary of

$$s_{X\cdots Y} = \frac{c_X s_X + c_{AB} s_{AB}}{c_X + c_{AB}} \qquad \text{(Eq. 10)}$$

which requires the concentration of co-sedimenting free $\mathbf{Y}$ of

$$\tilde{c}_Y = \frac{c_{AB}(s_{AB} - s_Y)}{Kc_Y(s_{AB} - s_Y) + (s_X - s_Y)} \qquad \text{(Eq. 11)}$$

such that the concentration of the undisturbed boundary is left as $c_{Y,tot} - \tilde{c}_Y - c_{AB}$. In this way, molecules of **A** and **B** sedimenting in the reaction boundary have the same time-average velocity $s_{X\cdots Y}$ -- a condition for a stable boundary.

Recognizing the sedimentation pattern of undisturbed and reaction boundaries, we may consider a signal-weighted average sedimentation coefficient of the system composed of these undisturbed and reaction boundaries:

$$s_w^{(EPT)} = \frac{\varepsilon_Y(c_{Y,tot} - c_{AB} - \tilde{c}_Y)s_Y + \left(\varepsilon_Y\tilde{c}_Y + \varepsilon_X c_X + (\varepsilon_X + \varepsilon_Y + \Delta\varepsilon)c_{AB}\right)s_{X\cdots Y}}{\varepsilon_Y c_{Y,tot} + \varepsilon_X c_{X,tot} + \Delta\varepsilon c_{AB}}$$ (Eq. 12)

After inserting the velocities and concentrations in the undisturbed and reaction boundary Eqs. 10 and 11, comparison with the traditional $s_w^0$ from Eq. 4, and after some rearrangements of terms we find

$$s_w^{(EPT)} = s_w^0 - \frac{\Delta\varepsilon\, c_{AB}}{\varepsilon_Y c_{Y,tot} + \varepsilon_X c_{X,tot} + \Delta\varepsilon c_{AB}} \times \frac{c_X}{c_X + c_{AB}} \times \left(s_{AB} - s_X\right)$$ (Eq. 13)

Thus, in the absence of signal changes $s_w^{(EPT)} = s_w^0$ and the population-averaged view (Eq. 4) and the EPT-based view (Eq. 12) are identical. Only in the presence of signal changes ($\Delta\varepsilon \neq 0$) is there a difference. Therefore, $s_w^{(EPT)}$ appears as a generalization of the population signal-weighted-average sedimentation coefficient, extending its applicability to rapidly reversible systems with signal changes in the complex.

To examine the physical meaning of the second term on the right hand side of Eq. 13, we consider a situation when only **X** contributes to the signal (**Y** is silent with $\varepsilon_Y = 0$, e.g., not carrying a fluorescent label) and assume the extreme case that **X** is completely quenched in the complex ($\Delta\varepsilon = -\varepsilon_X$): Now the additional term Eq. 13 reduces to $+\left(c_{AB}/(c_X + c_{AB})\right) \times \left(s_{AB} - s_X\right)$. We recognize the first factor as the fractional time **X** is in the complex and silent, which is multiplied by the second term that represents the additional migration that **X** will experience during this time. Therefore, the correction in the second term of Eq. 13 amounts to the migration that will be missed in the population based consideration of $s_w^0$.

*Determining Signal-Weighted Sedimentation Coefficients*

One convenient way to rigorously determine $s_w$ from given data comprising the evolution of signal boundaries is via fitting a diffusion-deconvoluted sedimentation coefficient distribution $c(s)$ to the raw data

$$a(r,t) \cong \int c(s)\chi_1(r,t,s,D(s))ds \qquad \text{(Eq. 14)}$$

where $\chi_1$ denotes a normalized Lamm equation solution (77, 78) for a species with sedimentation coefficient $s$ based on a hydrodynamic scaling law $D(s)$ (79). As previously shown elsewhere (5, 75), the temporal change in the radial integral in Eq. 3 defining the mass balance directly corresponds to the integral over the sedimentation coefficient distribution $c(s)$

$$s_w = \frac{\int c(s)sds}{\int c(s)ds} \qquad \text{(Eq. 15)}$$

for non-interacting mixtures. However, Eq. 15 also holds true more generally if a sufficiently good fit is achieved, in the sense that the integrals in Eq. 3 are faithfully matched. In practice, fits to interacting systems are usually excellent, due to the normal square-root dependent broadening of reaction boundaries (13). To this extent, despite the origin of $c(s)$ in the description of non-interacting systems it can be rigorously applied to data $a(r,t)$ of interacting systems (or any other systems) for the purpose of determining $s_w$. Essentially, $c(s)$ may be considered merely an efficient operational step to determine the integrals of Eq. 3. Similarly, $s_w$-values can be determined from apparent sedimentation coefficient distributions based on time-derivative analyses (5), but values will be skewed at finite time intervals (15). If the integration of $c(s)$ is carried out separately across the undisturbed and the reaction boundary, then amplitudes and $s_w$-values for both boundary components can be determined separately (67).

Beyond determination of $s_w$, integration of $c(s)$ will also provide a measure of the total signal of the loaded mixture, which can be used directly to assess deviations from linearity of signal as a function of loading concentration. Thus, this initial total signal can be used to observe and quantitate quenching or signal enhancement in titration series similar to measurements in a benchtop fluorometer.

The $c(s)$ distribution was calculated with the software SEDFIT (http://sedfitsedphat.nibib.nih.gov/software), and plots were created with the software GUSSI (80).

**Results and Discussion**

In theory, it is straightforward to account for altered quantum yields in the complexed fluorophore during the sedimentation experiment by using the solutions of the coupled Lamm equation (Eq. 1) to fit experimental data. In fact, methods for directly fitting the signal profiles of interacting systems with coupled Lamm equation are well-established and available in different current data analysis software (69, 81, 82). However, the exquisite sensitivity of sedimentation velocity to sample impurities and micro-heterogeneity of sedimentation and diffusion coefficients limits the applicability of this approach (67, 69). For binding partners that do not self-associate, a simple test is whether the individual components can be fitted well with a single-species Lamm equation solution, or significantly better with a distribution of species: if a single-species fit to an individual component fails to be satisfactory, then surely the interacting mixtures of these components should not be treated with discrete coupled Lamm equation solutions, either. In many or most studies, data fail this test due to extraneous signals from impurities and sample polydispersity. Another drawback of direct Lamm equation modeling is that it often over-parameterizes the description of boundary broadening, a fact that will be exacerbated by the addition of a $\Delta\varepsilon$ parameter.

Circumventing these difficulties, the most popular and robust approach is the initial description of the experimental data as diffusion-deconvoluted sedimentation coefficient distributions $c(s)$, which will usually provide excellent fits to the raw data (5, 75, 79). The superposition of $c(s)$ at different loading concentrations provides an overview of the system and lends itself to determine the binding mechanism and the time-scale of the reaction relative to the sedimentation. This can also resolve many impurities and allows us to exclude them from further consideration. In a second stage, the sedimentation coefficient distributions can be used as a platform to characterize the more robust features of the sedimentation pattern of interacting systems, such as the division between undisturbed and reaction boundary. Distributions can be integrated to extract their amplitudes and $s$-values, followed modeling of the resulting isotherms (67, 75, 83). Chiefly, this also includes the overall signal-weighted sedimentation coefficient $s_w$, encompassing transport from the entire interacting system. This has been a staple of sedimentation velocity analysis of interacting systems for many decades (2, 73).

Therefore, it will be important to understand and quantify the impact of changes in fluorescent quantum yield in complexes on the measured $s_w$. Our goal is not simply a partial differential equation, but a physical picture of how such signal changes impact the observed coupled transport. A key aspect of the sedimentation of rapidly reversible systems is that the reaction boundary contains fractions of both interacting components in their free state. This has been predicted in the seminal work of Gilbert and Jenkins in the 1950s (76) and subsequently experimentally verified. The population of co-sedimenting free species appears due to molecular co-transport while in complex, followed by transient release after finite complex lifetimes. In the present context, this basic phenomenon can be best demonstrated with simulated Lamm equation solutions for a hypothetical system with complete static quenching of signal in the complex. This is illustrated in Figure 2 for a 150 kDa, 7 S molecule **A** reversibly binding a silent 200 kDa, 9 S molecule **B** forming a silent 13 S complex, such that only free **A** contributes to the signal. If there was no co-sedimentation of **A** in the reaction boundary, then only a single 7 S boundary would appear. However, clearly a bimodal pattern arises with a fraction of the free molecules sedimenting in the faster reaction boundary.

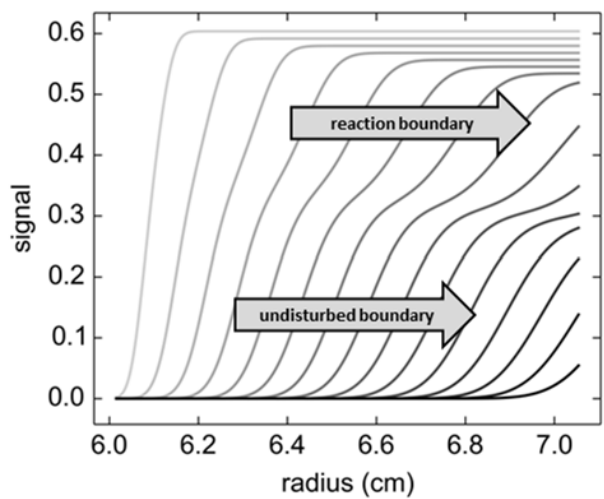

Figure 2: Sedimentation velocity profiles of a 150 kDa, 7 S component **A** sedimenting in the presence of an optically silent 200 kDa, 9 S component **B** forming a rapidly reversible complex with 13 S that complete quenches the signal, such that only free **A** generates a signal. Coupled Lamm equation solutions were calculated for loading concentrations $c_{A,tot} = c_{B,tot} = K_d$, in a solution column of 12 mm length at a rotor speed of 50,000 rpm; signal profiles are shown in 10 min intervals.

In the absence of quenching, the transport during the fractional time that **A** has spent in the complex will be accounted for in the population-averaged overall transport signal. In the presence of quenching, however, this co-transport is the origin of the failure of the conventional approach considering only equilibrium populations of species: Since only free **A** contributes to the signal, the conventional binding isotherm Eq. 4 would yield constant $s_w^0 = s_A$ throughout, unable to account for the extra co-transport of free **A** in the reaction boundary. A more productive view taken in EPT is the subdivision of **A** into a fraction that remains sedimenting free, and a fraction that exhibits coupled sedimentation with **B** such that their time-average velocity matches. In this view, the fraction of **A** in the reaction boundary also assumes a time-average signal, depending on the fractional time spent free or in complex. This picture reflects the physical mechanism of co-transport, and accounts for the physical transport each molecule experiences while silent in the complex.

In the following we test the implications of these considerations for $s_w$ isotherms with two model systems exhibiting altered fluorescence quantum yield of complexes. Using the coupled Lamm equation as a reference point, we carried out simulations with Eq. 1 at a range of concentrations, accounting for each species' signal contribution to generate perfect theoretical profiles of total signal as a function of time similar to those in Fig. 2. From these signal profiles, the sedimentation coefficients $s_w$ were extracted by fitting with $c(s)$ distributions followed by integration (Eq. 5). This is equivalent to the radial integration of Eq. 3 defining $s_w$, though more convenient and in correspondence with experimental practice (83). The resulting $s_w$ values can then be compared with the predictions from $s_w^{(EPT)}$ and $s_w^0$.

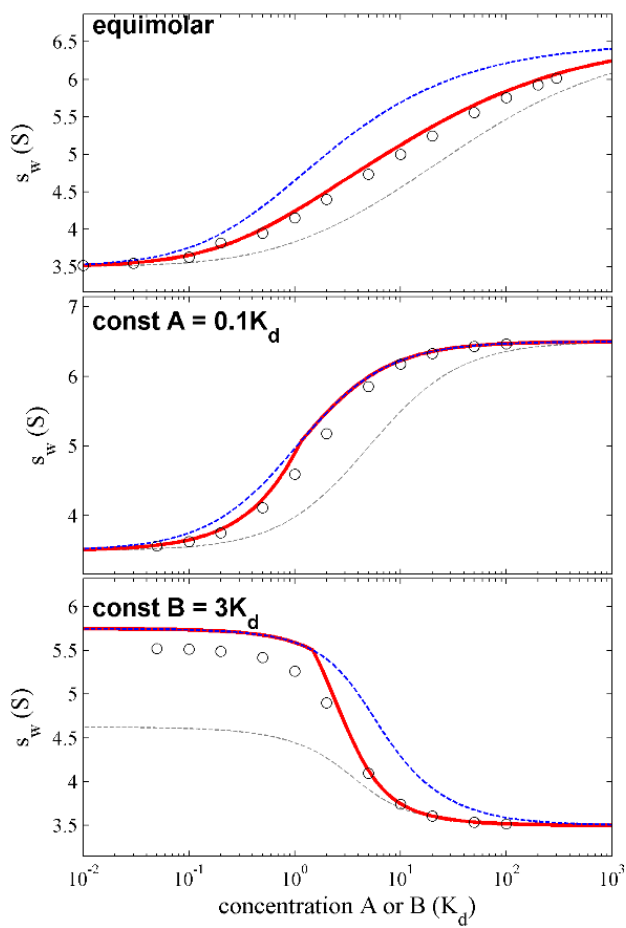

Figure 3: Comparison of $s_w$-values extracted from coupled Lamm equation solutions (Eq. 3/5, circles) with those from the standard population-based expression $s_w^0$ (Eq. 4, gray dotted line), and the isotherm from effective particle theory $s_w^{(EPT)}$ (Eq. 13, red solid line). Simulations were carried out for a system with a 45 kDa, 3.5 S component **A** sedimenting in the presence of an optically silent 75 kDa, 5 S component **B** forming a rapidly reversible complex with 6.5 S in a solution column of 12 mm length spinning at a rotor speed of 50,000 rpm. The signal of the complex was reduced to 0.2fold that of free **A**. For comparison, the $s_w^0$ isotherm in the absence of quenching is also shown (Eq. 4 with $\Delta\varepsilon = 0$, blue

dotted line). Concentrations were taken as equimolar dilution series (Top Panel), titration of constant concentration of **A** at 0.1fold $K_d$ (Middle Panel), and the titration of a constant concentration of **B** at 3fold $K_d$ (Lower Panel).

Results are shown in Fig. 3 for different concentration series for a system with signal only from component **A**, in the presence of quenching such that the complex signal is reduced to 0.2fold that of free form of **A**. It should be noted that there are no adjusted parameters in this plot. As may be expected, not accounting at all for signal quenching (blue dotted line) fails to describe the data (circles). In comparison, quenching causes the equimolar dilution series to exhibit a more shallow isotherm, and the titration series to be significantly steeper. When accounting for quenching of the complex simplistically without considering the coupled co-transport (gray dotted lines), the standard population-based isotherm $s_w^0$ performs even worse. The conventional isotherm $s_w^0$ significantly underestimates $s_w$ under all conditions and performs well only for conditions that lead to complete saturation or to the absence of any complex. Compared to both, the isotherm $s_w^{(EPT)}$ accounting for co-transport is much more consistent with the data throughout.

The discontinuity of the slopes in the predicted titration isotherms arise from the predicted transition points when the component that is completely engulfed in the reaction boundary switches. Where **A** is predicted by EPT to be engulfed entirely in the reaction boundary, mediated by an excess of **B**, the only detectable signal sediments with the $s$-value of the reaction boundary, and therefore signal quenching is predicted not to affect $s_w$. (This leads to the superimposed red and blue dashed traces in Fig. 3.) However, as can be easily discerned in the Lower Panel of Fig. 3 at low concentrations, these conditions also lead to the largest error in $s_w^{(EPT)}$: In this case, the concentration gradients from diffusion allow for a small population of free **A** to trail the reaction boundary, thereby lowering the observed $s_w$. Nevertheless, despite this error $s_w^{(EPT)}$ still performs much better than the conventional isotherm $s_w^0$ in this region, which would fail completely to recognize transport contributions from the complex in the reaction boundary. This is also the region where the difference between the isotherms is greatest, as shown in Fig. 4. With **B** in the range of $K_d$ and in excess over **A**, the largest fraction of **A** is subject to co-transport while maintaining a significant time-average signal from intermittent release in the free, signaling state.

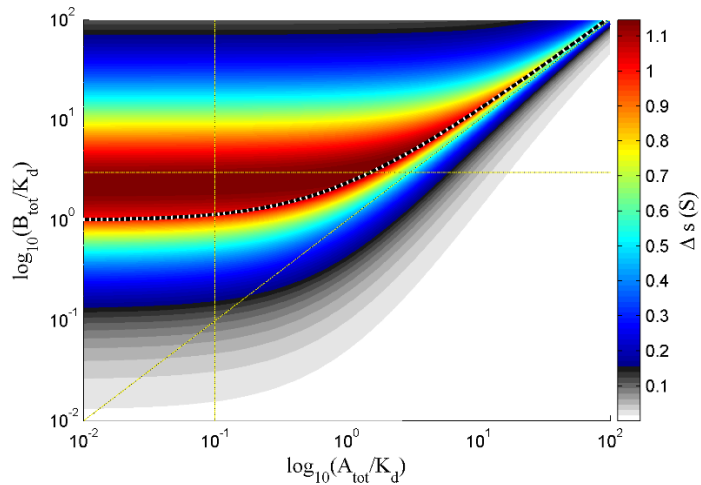

Figure 4: Overview of the difference between the quenching/co-transport isotherm from effective particle theory $S_w^{(EPT)}$ and the conventional population-based isotherm $S_w^0$ as a function of total loading concentration for the system of Fig. 3, encoded by color temperature. The phase transition line where the reaction boundary switches from completely engulfing all of **B** (lower right area) or all of **A** (upper left) is depicted as black-and-white dotted line. The thin yellow straight lines highlight the trajectories of the dilution and titration series employed in Fig. 3.

Finally, we examine the $s_w$ isotherms for the opposite case where the complex exhibits enhanced fluorescence. Fig. 5 shows the $s_w$ isotherms when complex signal is 3fold that of free form of **A** ( $\Delta\varepsilon = 2\varepsilon_A$, under otherwise identical conditions as in Fig. 3. As can be expected, the $s_w$-values are now significantly higher than in the case without signal change, though not as high as a simplistic population-based correction would predict. Again, the EPT-based prediction matches the sedimentation behavior of the system very well. (Even the case where **A** should be driven entirely into the reaction boundary is described very well, since the diffusionally escaped population of **A** now does not much contribute to the signal.)

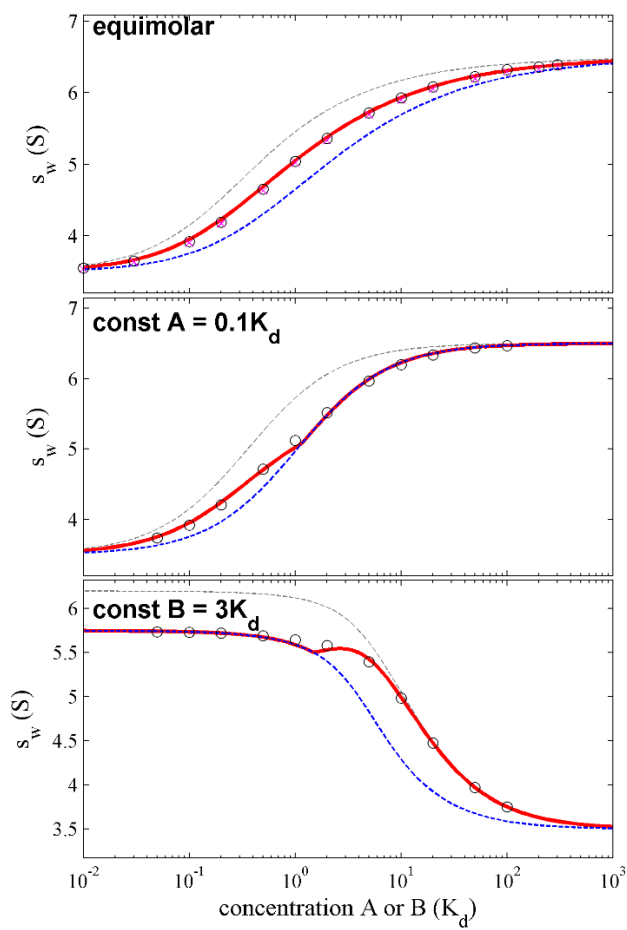

Figure 5: Impact of 3-fold fluorescence signal enhancement in the complex on the $s_w$ isotherms. Analogous to Fig. 3, and using the same molecular parameters, Lamm equation simulations were carried out to determine $s_w$ (circles), in comparison with ideal isotherm $S_w^0$ without signal change (blue dotted line), the standard population-based corrections to $S_w^0$ (gray dotted line), and the isotherm from effective particle theory $S_w^{(EPT)}$ (red solid line). As before, concentrations were taken as equimolar dilution series (Top Panel), titration of constant concentration of **A** at 0.1fold $K_d$ (Middle Panel), and the titration of a constant concentration of **B** at 3fold $K_d$ (Lower Panel).

## Conclusions

In summary, we have reported an anomaly in the observed sedimentation behavior for rapidly reversibly associating systems that exhibit changes of fluorescence quantum yield in the complex. In this case, the time-average state of all molecules is the population average state, but the time-average signal transport is not equal to the population-average signal transport. It leads to significant deviations from the conventionally expected isotherms of $s_w$ based on equilibrium populations. This effect is another incarnation of the seemingly counter-intuitive properties of the reaction boundaries if viewed in a naive species-based picture without considering the mechanism of coupled co-transport. Like other `anomalies', the effective particle theory naturally explains these features, as it reflects the physical reality of a coupled migrating system. A more general isotherm of $s_w$ based on EPT provides a satisfactory quantitative approximation of this effect.

Nothing in the derivations considerations refers in any way to the nature or sign of the signal changes associated with complex formation.  Therefore, the same will hold for static or dynamic quenching, fluorescence enhancement, or any other photophysical process altering the fluorescence quantum yield.  It should likewise apply to absorbance data when hyper- or hypochromicity induced by binding events alters extinction coefficients, which may be observed in nucleic acid interactions or protein interactions involving large conformational changes.  One very promising area of future developments involves the combination of mass based separation in sedimentation with signals that offer proximity-dependent information on binding, such as Förster resonance energy transfer.  This may potentially enable experiments to transcend the limitations of macroscopic binding studies in SV.  Based on the results of the present work, analyses of such experiments must account for the true physical nature of the coupled co-sedimentation.

## Author Contributions

S.K.C. performed research; analyzed data; H.Z. designed research; wrote the manuscript; P.S. analyzed data; designed research; wrote the manuscript.

## Acknowledgment

We thank the anonymous reviewers for excellent suggestions.  This work was supported by the Intramural Research Program of the National Institute of Biomedical Imaging and Bioengineering, National Institutes of Health.